\pdfoutput=1
\documentclass[10pt,a4paper]{article}
\usepackage{times}
\usepackage[active]{srcltx}
\usepackage{graphics}
\usepackage{graphicx}
\begin{document}
\large
\date{ }

\begin{center}
{\Large\bf Correlation Time-of-flight Spectrometry of Ultracold Neutrons}

\vskip 0.5cm

M. I. Novopoltsev$^{a}$ and Yu. N. Pokotilovski$^{b}$ \footnote{e-mail: pokot@nf.jinr.ru}

\vskip 0.5cm

$^{a}$Mordovian University, Saransk, 430005 Republic of Mordovia, Russia

$^{b}$Joint Institute for Nuclear Research, Dubna, Moscow Region, 141980 Russia

\vskip 0.5cm

{\bf Abstract}

\begin{minipage}{130mm}

\vskip 0.7cm

 The features of the correlation method used in time-of-flight spectrometry of
ultracold neutrons are analyzed.
 The time-of-flight spectrometer for the energy range of ultracold neutrons is
described, and results of its testing by measuring spectra of neutrons passing
through interference filters are presented.
\end{minipage}
\end{center}
\vskip 0.3cm

Ultracold neutrons (UCN) with energies below $\sim 0.25 \mu$eV [1] can be
confined for a long period of time within a closed volume.
 This phenomenon has found applications in investigations of the fundamental
properties of a neutron [2].

 Owing to a very low energy of these neutrons, high energy ($\sim$ 1 neV) and
momentum ($\sim 10^{-4} \AA^{-1}$) resolutions can basically be attained in the
UCN scattering experiments.
 However, small number of experiments have been performed in this energy range.
 The modest scale of these investigations is explained by the low
intensity of neutron fluxes at these energies.
 The best neutron source in this energy range, which is located at the Institut
Laue--Langevin [3], yields $\sim$ 50 neutrons/(cm$^{2}$ s neV) in the maximum
of the neutron spectrum near 750 neV (v$\sim$12 m/s), and the neutron density
in the storage mode is $\sim$40 cm$^{-3}$.
 It is now expected that, after commissioning of the new UCN sources [4--6],
considerable progress will be achieved in the intensity of neutron fluxes with
very low energies owing to the use of higher-efficiency cold moderators---solid
deuterium or solid deuterium hydrocarbons at temperatures of 5--10 K -- for
generation of very cold neutrons [7].

 However, at available low intensities, the statistical accuracy of
measurements is improved using special methods, e.g., the correlation method
discussed in this paper.

\vspace{2mm}

CORRELATION METHOD IN TIME-OF-FLIGHT SPECTROMETRY

 The significant drawback of the standard time-of-flight (TOF) method is its
low luminosity, which is defined by the ratio of the duration of the neutron
pulse produced by the chopper to the modulation period of the particle beam.
 Numerous available neutron TOF spectrometers have a duty cycle of a few
percent or under.
 In 1964, the authors of [8, 9] independently proposed using the correlation
method, which was formerly employed in various fields of experimental physics,
in TOF spectrometry.
 An impetus for further development of the correlation method was received in
designing neutron and molecular beam spectrometers (see reviews in [10--12]).

 At the initial stages of investigations with UCNs, the neutron fluxes were
extremely low, and, sometimes, the neutron background under the experimental
conditions was high.
 This required that the correlation method be used in UCN TOF spectrometry,
since it offers a chance to overcome these difficulties.
 Let us briefly consider the basic principles of the correlation TOF
spectrometry.
 This method implies that a chopper opens $n$ times during beam modulation
period $T$ rather than once, as is the case of the classical TOF method.
 As a result, the effective particle flux incident on the detector increases
$n$ times, whereas the background level remains constant.
 The chopper modulates the particle beam according to the law
\begin{equation}
S(t)=\sum_{i=0}^{N-1}a_{i}g(t-i\Delta T),
\end{equation}
where $g(t)$ describes beam modulation by a single-slit chopper, and $N$ is the
number of elementary intervals with duration $\Delta T$ in the period of
sequence $T=N\Delta T$.
For the standard chopper,
\begin{equation}
a_{i}=\Biggl\{{1 \quad \mbox{ if } i=0 \atop  0 \quad \mbox{ if } i=1,2,...,N-1,}
\end{equation}
and, for the correlation method, $a_{i}$ is the binary sequence of zeros and ones.
 If channel width of the time analyzer $\delta\tau$ is equal to elementary
interval $\Delta T$, the number of neutrons detected in the $k$-th channel is
\begin{equation}
z_{k}=\sum_{i=0}^{N-1}a_{k-i}f_{i}+u_{k},
\end{equation}
where $f_{i}$ is the TOF spectrum corresponding to the $i$-th channel of the
analyzer, and $u_{k}$ is the mean background level in the channel.
 In the matrix form, this equation is
\begin{equation}
Z=A\cdot F+U,
\end{equation}
where $Z$ is the $N$-dimensional vector, the elements of which are counts
$z_{k}$ in the analyzer channels; $F$ and $U$ are the vectors of the TOF spectrum
and background, respectively; and $A$ is the cyclic matrix $N\times N$, the
rows of which have been obtained by cyclic permutation of sequence $a_{i}$.
 For the standard chopper, matrix $A$ transforms into unit matrix $I$.
 From the last two expressions, it follows that the TOF spectrum
(at $det A\not=0$) is
\begin{equation}
F=B(Z-U),
\end{equation}
where
\begin{equation}
B\cdot A=I.
\end{equation}
 If the width of all analyzer channels is uniform, the background is
independent of the channel number (i.e., $u_{i}=u$), therefore,
\begin{equation}
f_{k}=\sum_{i=0}^{N-1}b_{i-k}z_{i}-u\sum_{i=0}^{N-1}b_{i-k}.
\end{equation}
 Equation (6) leads to the system of linear equations
\begin{equation}
\sum_{i=0}^{N-1}b_{i}a_{i-k}=\Biggl\{{1 \quad \mbox{ if } k=0 \atop 0 \quad \mbox{ if } k=1,2,...,N-1,}
\end{equation}
 Solving these equations, we determine the $b_{i}$ values.
 Therefore, Eqs. (5) and (7) allow us to obtain TOF distribution $F$ from
distribution $Z$ recorded by the analyzer.
 The variance of distribution $F$ can be determined from Eq. (7) using the
Poisson distribution,
\begin{equation}
\sigma^{2}(f_{k})=\sum_{i=0}^{N-1}b^{2}_{i-k}z_{i}+u\sum_{i=0}^{N-1}b^{2}_{i-k}.
\end{equation}
 Binary sequence $a_{i}$ is selected so that function (9) has an absolute
minimum.
 In the general form, this problem was solved in [13] using the procedures of
the theory of optimization; the classes of quasi-random binary sequences were
found, so that they satisfied condition (9)
\begin{equation}
\sum_{i=0}^{N-1}a_{i}a_{i+j}=\Biggl\{{n \quad \mbox{ if } j=0 \atop m=\frac{n(n-1)}{N-1} \quad \mbox{ if }
j\not=0,}
\end{equation}
where $m$ is an integer, and $n=\sum_{i=0}^{N-1}a_{i}$ is the number of unities
in sequence $a_{i}$ and, therefore, the number of open windows in the correlation
chopper.
 Quantity $c=(n-1)/(N-1)$ characterizes the operational cycle of this sequence.
From Eqs. (7) and (8), it follows that, according to [14],
\begin{equation}
f_{k}=\frac{N-1}{n(N-n)}\sum_{i=0}^{N-1}a_{i-k}z_{i}
-\frac{n-1}{n(N-n)}\sum_{i=0}^{N-1}z_{i-k}-\frac{u}{n}.
\end{equation}
 This formula shows that the time-independent background decreases $n$ times.
 Apart from the useful signal presented by the first term in the right part of
Eq. (11), the information in the $k$-th channel contains an additional
background term (the second term) proportional to the total useful signal.
 Gain $\gamma$ of the correlation TOF method in comparison with the conventional
method is equal to the ratio of the measurement times needed in either method
to ensure the same statistical accuracy and the same resolution:
$\gamma=\sigma^{2}_{st}/\sigma^{2}_{corr}$.
 At $\gamma >1$, the correlation chopper is more advantageous.
 Pseudorandom maximum-length sequences [15] with $N=2^{p}-1$ ($p$ is an integer,
$n=(N+1)/2=2^{p-1}$ è $c=1/2$) are most frequently used in the correlation
method to modulate the particle flux.
 Therefore,
\begin{equation}
b_{i}=\Biggl\{{1/n \quad \mbox{ if } a_{i}=1 \atop -1/n \quad \mbox{ if } a_{i}=0,}
\end{equation}
 In [9], it was shown that, in practice, for the true $f_{k}$ spectrum to be
reconstructed, it was profitable to use pseudorandom binary sequence
$a_{i}^{'}=2a_{i}-1\equiv\pm 1$, satisfying the conditions
\begin{equation}
\sum_{i=0}^{N-1}a^{'}_{i}a^{'}_{i+k}=\Biggl\{{N \quad \mbox{ if } k=1 \atop -1 \quad \mbox{ if } k\not=0,} \quad
\quad \sum_{i=0}^{N-1}a^{'}_{i}=1.
\end{equation}
 In this case, the second term disappears in Eq. (11), and
\begin{equation}
f_{k}=\sum_{i=0}^{N-1}a^{'}_{k-i}z_{i}-u.
\end{equation}
 From Eq. (14), it follows that data $z_{i}$ stored in all the analyzer channels
are used to obtain an individual point in spectrum $f_{k}$.
 Therefore, variance $\sigma^{2}(f_{k})$ for the spectrum under investigation
will be the same for all spectrum points, being determined by the sum of counts
in all channels
\begin{equation}
\sigma^{2}(f_{k})=\sum_{i=0}^{N-1}z_{i}+\sigma^{2}(u).
\end{equation}
 For the $k$-th analyzer channel, the gain is [14]
\begin{equation}
\gamma_{k}=\frac{n}{N}\frac{f_{k}+u_{k}+\sigma^{2}_{k}} {\bar
f+\frac{u_{k}+\sigma^{2}(u_{k})}{n}},
\end{equation}
where $\bar f=\frac{1}{N}\sum_{i=0}^{N-1}f_{i}$ is the mean value of the TOF
spectrum and $\sigma^{2}(u_{k})$ is the variance of the background measurement
in the $k$-th channel.

 In measurements of the UCN spectra, the background is independent of the
channel number (stationary neutron sources are used), and the condition
$\sigma^{2}(u_{k})\ll u_{k}$ is usually satisfied, since averaging of the
background reduces its variance.
 From Eq. (16), we obtain that
\begin{equation}
\gamma_{k}=\frac{n}{N}\frac{f_{k}+u} {\bar f+\frac{u}{n}}.
\end{equation}
 As a result, the gain depends on the shape and total area of the distribution,
the background level, and relative transmission of the chopper $n/N$.
 At a low background, the use of the correlation analysis is advantageous for
those parts of the spectrum where the sum of the useful signal and the
background ($f_{k}+u$) is at least twofold higher than the mean counting over
the whole recorded spectrum.
 In addition, the correlation method is undoubtedly advantageous at a
suppressing background.

\vspace{2mm}

DESIGN OF THE UCN CORRELATION SPECTROMETER

 Two spectrometers -- small and large -- were produced to measure UCN spectra;
they had 42- and 160-cm-diameter choppers, respectively.

 The schematic diagram of the larger spectrometer is shown in Fig. 1.
 The chopper of this spectrometer, which is enclosed in vacuum housing {\it 2},
comprises stainless steel rim {\it 3} (diam. 160 cm).
 Spokes {\it 4} join it to bushing {\it 8} that revolves on immobile hollow axle {\it 10}.
 The spokes have made it possible to reduce the chopper weight; they are also
convenient in adjustment of axial and radial run-out.

 The chopper has 63 metal plates {\it 1}, which are attached to the rim and are
used to modulate the neutron flux according to the pseudorandom sequence with
$N=127$.
 These plates, shaped as a trapezoid with bases of 39.4 and 42.2 mm and a
height of 60 mm, were produced from 0.1-mm-thick bronze foil according to a
special template with an accuracy of 0.05 mm.
 To flatten the surface and impart elastic properties to the plates, they were
squeezed between two flat steel disks and subjected to special thermal
treatment.
 In addition, 6-$\mu m$-thick cadmium layers were electrochemically deposited
on both sides of each chopper plate.

 Special emphasis was placed on the accuracy with which the chopper plates were
fixed in place on the rim, since (as will be shown in what follows) the spread
in the size of the windows in the chopper sequence could lead to appreciable
distortions of the spectrometer resolution function and to fluctuations in the
reconstructed TOF spectrum.

 The driving torque is transmitted from AOL-22-4-type three-phase asynchronous
motor {\it 15} via step-down reduction gear {\it 14} and magnetic clutch {\it 12} into the
vacuum to chopper shaft {\it 11}.
 The internal and external parts of the magnetic clutch are separated by a
stainless steel membrane  {\it 13} 0.2 mm thick.

 The start pulses from the chopper, which indicate the beginning of the
modulation cycle, are produced by means of electric light bulb {\it 6}, photodiode
{\it 7}, and mirror {\it 9}.
 The chopper's period of revolution can be varied within the range of 123--127
ms by changing the supply voltage of the motor.

 There is branch pipe {\it 20} with an inner diameter of 100 mm and a flange for
connecting the spectrometer to the setup under investigation, as well as flange
{\it 16} for connecting detector {\it 19}.
 Branch pipe {\it 20} contains a collimator, samples under investigation, or a
mirror neutron guide tube that guides neutrons to the chopper.
 The entrance window has variable dimensions, which are conventionally
40$\times$40 mm$^{2}$.

 Past the chopper, there are replaceable rectangular mirror neutron guide tubes {\it 18}
with an inner cross section of 45$\times$50 mm$^{2}$ and different lengths.
 The neutron guide tubes are made from FLOAT glass bands with 0.2-$\mu m$-thick
nickel layers sputtered onto their surface.
 The characteristics of the neutron guide tubes produced thereby and used to
extract thermal and cold neutrons from a reactor were described in [16, 17].
 The presence of a neutron guide tube in the flight path inevitably results in a
fraction of neutrons being nonspecularly reflected from the surface, thus
distorting the measured spectrum.
 Therefore, short neutron guide tubes (20--50 cm long) were used in the
majority of measurements.
 If the neutron guide tube is 50 cm long, the elementary interval in the
modulating sequence is 1 ms, and the neutron energy is 100 neV, the energy
resolution is 2 neV.

 The other important factor in UCN TOF spectrometry is the reflection of
neutrons from the detector surface and accumulation of neutrons vagabonding
between the detector and the chopper in the neutron guide tube.
 This also leads to distortions in the spectrum.
 At the end of flight path {\it 17}, there is an "all-wave" UCN detector, described in [18], with
a rotating disk 420 mm in diameter with a corrugated paraterphenyl scintillator
and a $^{6}$LiF radiator 700 $\mu$g/cm$^{2}$ thick evaporated onto it.
 Such a detector does not reflect neutrons [18].

 The spectrometer was pumped via branch pipe {\it 5} by an independent system
consisting of a backing pump, three absorption zeolite pumps, and a NORD-250
ion pump.
 During the measurements, the vacuum inside the spectrometer was
$\sim 5\times 10^{-4}$ Pa.
 The spectrometer and the pumping system are placed on a cart, which is capable
of moving the spectrometer over the experimental area and in a vertical plane.

 The detector signals arrived at the input of the system of data acquisition
and processing, which was made to the CAMAC standard [19].
 Operation of this system was controlled from a microprocessor-based crate controller.
 Following the preset program, the system performed acquisition of primary
spectra, calculation of the correlation function according to Eq. (14), and
typing out of the results.
 The 127-channel time-to-digital converter was triggered once in a cycle by the
start pulses from the chopper.
 These pulses were used to measure the revolution period of the chopper and
monitor its stability.

 A small spectrometer with an entrance window of 5$\times$50 mm$^{2}$ was used
to develop and test the experimental procedure.
 Its electric drive was similar to the above-described one.
 The glass neutron guide tubes had an inner cross section of
25$\times$50 mm$^{2}$.

\vspace{2mm}

RESOLUTION FUNCTION OF THE UCN CORRELATION SPECTROMETER

 The resolution function of the correlation spectrometer is defined in [9] as
\begin{equation}
R_{k}=\sum_{i=0}^{N-1}s_{i}\cdot a^{'}_{j-k},
\end{equation}
where
\begin{equation}
s_{i}=\int_{(i-1)\Delta T}^{i\cdot\Delta T}s(t)dt.
\end{equation}
 $s(t)$ is the transmission function of the chopper for a UCN flux, and
$a_{i}^{'}$ is the pseudorandom binary sequence of numbers that satisfies
conditions (14) and is used to reconstruct TOF spectrum $f_{k}$.

 If the spectrometer's entrance window is shaped as a narrow radial slit,
$s(t)$ is presented by a sequence of rectangles with elementary interval
$\Delta T$, and the resolution function is shaped as a rectangular peak with
width $\Delta T$ on a zero pedestal.
 When the beam has a width of the chopper's elementary interval, $s(t)$ is
presented by alternating triangles and trapezoids, which causes the resolution
to deteriorate twofold [14].

  The resolution function of the actual correlation TOF spectrometer may differ
from the ideal case for a number of reasons.
 This is usually displayed by a deformation of its main peak, appearance of
ghost peaks, and modulation of the remaining spectrum.
 In our case, this results, e.g., from the instability of the chopper's period
of revolution, the finite thickness of its plates, and the inaccuracy in their
production and mounting on the disk.

 Let us consider the effect of each factor on the spectrometer resolution
function.
 The inevitable instability of the chopper speed results in a phase shift
between the analyzer scan and the modulating sequence.
 In the classical TOF spectrometer, this leads only to deterioration of the
resolution and, in the correlation spectrometer, gives rise to ghost peaks.
 This fact is particularly important for the described spectrometer design,
since, due to the presence of the magnetic clutch, stabilization of the motor
speed fails to ensure the appropriate stability of the chopper speed.
 Therefore, the time-to-digital converter of the time analyzer is equipped with
a system for tuning the analyzer channel width (and, hence, its scanning period)
to the chopper's period of revolution.

 The resolution function was calculated for the case of the worst mismatch over
the period (0.1 of the channel width).
 The results are presented in Fig. 2a.
 The calculated mismatch gives rise to additional counts in all analyzer
channels, which do not exceed in each channel a value of 1.4\% of the main peak
area.
 In addition, a 5\% fraction of the peak intensity was swapped from the last
channel to the first; however, this swap affected the measured spectra only
slightly.
 In reality, the mismatch of the chopper and analyzer periods is considerably
lower than the value used in the calculation (this mismatch was checked in the
measurements).

 The workmanship of the chopper is the other factor affecting the resolution
function.
 It results in a spread in the width of the windows in comparison with the
ideal case and distorts the resolution function.
 It is reasonable to assume that, when the chopper is manufactured, the boundary
of each window randomly departs from the ideal position.

 Figure 2b presents the calculated histogram of the resolution function for the
random uniform scatter of the chopper elements' boundaries within the limits of
$\pm$0.5 mm (1.25
 A total of 34 variants of random departures in the window widths were
calculated, and none of them yielded an anomaly with a substantially increased
variance of counts in the channels.
 The histogram in Fig. 2b is typical: the amplitude of new peaks is 0.5\% of
the main peak or less.
 At a narrower chopper's elementary interval (e.g., 5 mm for the reduced
spectrometer model), the relative scatter of the window boundaries may
basically be larger.
 Figure 2c presents the resolution function at a uniform scatter of the window
boundaries, which is 12.5\% of the width of the elementary interval.
 In this case, the amplitude of the peaks does not exceed 3\% of the main peak.

 The thickness of the chopper has an appreciable effect on the shape of the
spectrometer's resolution function in spectrometry of very cold neutrons.
 At chopper thickness $d$, width $h$ of the slit in it, and chopper's linear
velocity $w$, the relative transmission of the chopper window depends on
neutron velocity $v$ and is defined as
\begin{equation}
\varphi(v)=\frac{\Phi_{w}}{\Phi_{0}}=1-\frac{wd}{vh}\alpha, \quad (wd\ll vh),
\end{equation}
where $\Phi_{0}$ and $\Phi_{w}$ are the fluxes of neutrons passing through a
single window when the chopper is at rest and moves with velocity $w$,
respectively.

 Coefficient $\alpha$ depends on the type of the sequence element next to the
given window, it is equal to zero or unity if the next element is a slit or a
crossbar, respectively.
 Since, the chopper has only one window in the case of standard TOF
spectrometry, $\alpha$=1 and spectrum distortion is reduced only to its
multiplication by term $\varphi (v)$, which is not very important in relative
measurements.
 In the correlation chopper, there are $(N + 1)/2$ windows located irregularly
(pseudorandomly); therefore, $\alpha$ can take on values 0 and 1, which causes
deformation of the resolution function.

 The distortions in the spectra produced thereby were considered in [14].
 For the described correlation spectrometer, the TOF spectra of monoenergetic
neutrons with velocities of 2, 3, 4, and 6 m/s were numerically simulated at a
flight path of 20 cm, $\Delta\tau=1$ ms, and chopper thicknesses of 0.1, 0.2,
and 0.8 mm.
 From histograms shown in Fig. 3, it is apparent that each spectrum peak is
deformed and accompanied by two groups of false components.
 One of them is located on each side of the main peak and worsens the
spectrometer resolution only slightly.
 The other, which has a negative amplitude, is located 8 channels to the left,
and causes distortions in the spectrum.

 Since the peak amplitude increases significantly with an increase in the
chopper thickness and a decrease in the neutron velocity, it is apparent that
the chopper plates must have the minimum thickness.
 In our case (the plate thickness is 0.1 mm) with the neutron velocity as low as
1 m/s, the negative spike makes 5\% of the main peak.
 At a known dependence of the chopper transmittance on neutron velocity $v$,
there exists a correcting algorithm for taking into account the effect of the
chopper thickness. For $N=127$, it assumes the form [14]
\begin{equation}
f_{k}^{'}=f_{k}+\epsilon(v)\cdot f_{k-119},\quad  k=0,1,1,...,126,
\end{equation}
where $\epsilon(v)$ is the term dependent on the chopper thickness, its linear
velocity, and neutron velocity $v$; $f_{k}$ and $f_{k}^{'}$ are the measured
and corrected spectra, respectively.

 Actual resolution function $R_{k}$ and modulation of particle flux $s^{*}_{j}$
were experimentally determined using an alpha-source placed in front of the
chopper and simulating neutrons with an infinite velocity.
 Such measurements took into account all above-described distortions of the
resolution function, except for the influence of the chopper thickness.
 The histograms obtained for the resolution function of the spectrometer and
its reduced model are presented in Figs. 4a and 4b, respectively.

 It is apparent that the maximum deviation of $R_{k}$ from the zero level does
not exceed 1 and 3\% of the main peak, respectively.
 The results of our simulations and measurements, as well as the preliminary
analysis according to (15)--(17), show that the correlation method is generally
disadvantageous in studying weak effects against background of stronger ones.
 It is applicable in investigations of intense peaks, as well as in spectrum
measurements at a low neutron flux intensity and a relatively high level of
noncorrelated background, which are exactly the conditions for which the
spectrometer has been designed.

\vspace{2mm}

UCN TRANSMISSION THROUGH INTERFERENCE FILTERS

 In order to test the spectrometer efficiency and experimentally ascertain the
advantages of the correlation TOF spectrometry method, we measured the UCN
transmission through interference filters, with which it was possible to form
narrow resonance lines in the neutron spectrum.

 The interference effects, which are well-known optical phenomena, appear in
layered systems and manifest themselves in the resonance character of the
dependence of the light-transmission factor on the wavelength.
 It is apparent that similar effects can be observed in neutron optics.

 Interference filters for neutrons were described in [20, 21].
 The filter in [21] was based on the well-known quantum-mechanical example of
particle's transmission through a system of two potential humps [22].
 In this case, quasi-stationary states corresponding to the resonance character
of UCN transmission through a filter composed of alternating thin layers of
substances with different heights of the potential barrier for neutrons may
exist in the space between the humps.
 The operation of such a filter was described for the first time in [23].
 A gravity diffractometer [24] was used in this study for spectrometry
measurements.

 Our filters were produced by sequential vacuum deposition of copper and
aluminum layers on polished silicon wafers.
 The spectra of UCNs passed through the filter are shown in Figs. 5a and 5b.
 Using the least squares method, we determined that, with due account of the
spectrometer resolution function, the best spectrum representation was attained
at the Gaussian distribution of the middle layer thicknesses with variances of
42 and 28 \AA.
 The widths of the resonance levels were 8, 15, and 18 neV for levels with
energies of 107, 122, and 148 neV, respectively.

 The measured transmission of the interference filters has demonstrated the
advantages of the correlation TOF spectrometer over the standard one and the
other spectrometer types in recording spectra of very slow neutrons with a
resonance structure.
 Ten days were spent in [23] on measurement of a single spectrum with the
gravity diffractometer [24] at an incident UCN intensity of
$\sim$5 neutrons/(cm$^{2}$ s) (which was close to the UCN flux in our study)
and a filter area of 200 cm$^{2}$; in this case, the statistical accuracy
obtained in the peak was 10\%.
 The time it took to attain the same accuracy in our measurements was 1/20 of
this value when the small spectrometer and a 2.5-cm$^{2}$ filter were used and
only 2 h for the larger chopper and a 4$\times$4 cm$^{2}$ filter.
 Therefore, the efficiency of our spectrometer was at least two orders of
magnitude higher than that of the spectrometer in [24].

 On the other hand, there is a significant gain in comparison with the standard
TOF method: according to Eq. (17), the correlation method has benefits
practically in all points of the peaks under conditions of a low background
level (at a signal-to-noise ratio of $\sim$4).
 The gain in the data acquisition time at a fixed statistical accuracy in the
peak maxima was 5--10 and even greater when the background exceeded the useful
signal by a factor of 4.

 The described here spectrometers were constructed [25] at the early time of
investigations in the UCN field, when neutron fluxes were very low -- several
neutrons/(cm$^{2}$ s.
 They have been used in a number of experiments:
in the measurement of the UCN absorption coefficient at subbarrier reflection
from the potential wall [26], in the demonstration of "metallic reflection"
of neutrons from the surface with large absorption cross secton [27].

 The correlation type spectrometer with thin-film ferromagnetic chopper [28] is
described in [29].

\vspace{2mm}

REFERENCES

[1] Shapiro, F.L., {\it Rroc. Intern. Conf. "Nuclear Structure with Neutrons,}"
Budapest, 1972; Ero, J. and Szucs, J., Eds., New York: Plenum, 1972, p. 259;

Steyerl, A., Springer Tracts in Modern Physics, Berlin, Heidelberg, New York:
Springer, 1977, vol. 80, p. 57;

Golub R. and Pendlebury J.M., Rep. Progr. Phys. 1979, vol. 42, p. 439;

Ignatovich V.K, {\it Fizika ul'trakholodnykh neitronov}, Moscow: Nauka, 1986.

{\it The Physics of Ultracold Neutrons}, Oxford: Clarendon, 1990;

Golub R., Richardson D.J., Lamoreaux S., {\it Ultracold Neutrons}, Bristol:
Adam Hilger, 1991;

Pendlebury J.M., Ann. Revs. Nucl. Part. Sci., 1993, vol. 43, p. 687.

[2] {\it Proc. Intern. Conf. "Fundamental Physics with Slow Neutrons,}"
Grenoble, France, 1998, Nucl. Instrum. Methods Phys. Res. A, 2000, vol. 440;

{\it Proc. Intern. Conf. "Fundamental Physics with Slow Neutrons,}"
Gaithersburg, USA, 2004, J. Res. NIST, 2005, vol. 110, nos. 3, 4;

{\it Proc. Intern. Workshop "Particle Physics with Slow Neutrons,}" ILL,
Grenoble, France, 2008, Nucl. Instrum. Methods Phys. Res. A, 2009, vol. 611,
nos. 2--3.

[3]. Steyerl, A., Nagel, H., Schreiber, F.-X., et al., Phys. Lett. A., 1986,
vol. 116, p. 347.

[4] Hill, P.E., Anaya, J.M., Bowles, T.J., et al., Nucl. Instrum. Methods Phys.
Res. A, 2000, vol. 440, p. 674.

[5] Trinks, U., Hartmann, F.J., Paul, S., and Schott, W., Nucl. Instrum.
Methods Phys. Res. A, 2000, vol. 440, p. 666.

[6] {\it Proc. Workshop on PSI UCN Source}, PSI, Villigen, Switzerland, 2001;

http://ucn.web.psi.ch/techrev\_ucn/htm

[7] Golub, R. and B\"oning, K., Z. Phys. B, 1983, vol. 51, p. 95;

Yu, Z. Ch., Malik, S.S., Golub R. Z. Phys. B 1986, vol. 62, p. 137;

Serebrov, A.P., Mityukhlyaev, V.A., Zakharov, A.A., et al., Pis'ma Zh. Eksp.
Teor. Fiz., 1994, vol. 59, p. 728 [JETP Lett., vol. 59, p. 757];

Serebrov, A.P., Mityukhlyaev, V.A., Zakharov, A.A., et al., Pis'ma Zh. Eksp.
Teor. Fiz., 1995, vol. 62, p. 764 [JETP Lett., vol. 62, p. 785];

Serebrov, A.P. Nucl. Instrum. Methods Phys. Res. A, 2000, vol. 440, p. 653;

Serebrov, A., Mityukhlyaev, A, Zakharov, A., et al, Nucl. Instrum. Methods
Phys. Res. A, 2000, vol. 440, p. 658.

[8] Cook-Yahrborough, H.E., {\it Instrumentation Techniques in Nuclear Pulse
Analysis}, Washington, DC, 1964.

[9] Mogil'ner, A.I., Sal'nikov, O.A., and Timokhin, L.A., Prib. Tekh. Eksp.,
1966, no. 2, p. 22.

[10] Gl\"azer, V., Fiz. Elem. Chastits At. Yadra, 1972, vol. 2, p. 1125.

[11] Kroo, N. and Cher, L., Fiz. Elem. Chastits At. Yadra, 1977, vol. 8,
p. 1412.

[12] Tsitovich, A.P., Prib. Tekh. Eksp., 1976, no. 1, p. 7.

[13] Hossfeld, S. and Amadori, L., Julich Rep./JUL-684-FF, 1970.

[14] Buevoz, J.L. and Roult, G., Rev. Phys. Appl., 1970, vol. 12, p. 591.

[15] Peterson, W.W., {\it Error Correcting Codes}, Cambridge, MA: MIT Press,
1961.

[16] Kornilov, D.V., Nazarov, V.M., Sysoev, V.P., et al., JINR Communic.
P13-80-496, Dubna, 1980.

[17] Kashukeev, N.T. and Chikov, N.F., JINR Preprint P3-82-145, Dubna, 1982.

[18] Novopoltsev, M.I. and Pokotilovskii, Yu.N., Nucl. Instrum. Methods Phys.
Res. A, 1980, vol. 171, p. 497.

[19] Barabash, I.P., Gubarev, E.Yu., Elizarov, O.I., et al.,
JINR Communic. 11-12433, Dubna, 1979.

[20] Antonov, A.V., Isakov, A.I., Kazarnovskii, M.V., et al., Pis'ma Zh. Eksp.
Teor. Fiz., 1974, vol. 20, p. 32.

[21] Seregin, A A., Zh. Eksp. Teor. Fiz., 1977, vol. 73, p. 1634.

[22] Bohm, D., {\it Quantum Theory}, Englewood Cliffs, New Jersey:
Prentice-Hall, 1951, Chapter 12.

[23] Steinhauser, K.-A., Steyerl, A., Schekenhofer, H., et al.,
Phys. Rev. Lett., 1980, vol. 44, p. 1306.

[24] Scheckenhofer, H. and Steyerl, A., Phys. Rev. Lett., 1977, vol. 39,
p. 1310.

[25] M.I. Novopoltsev, Yu.N. Pokotilovskii, "Correlation spectrometer for
ultracold neutrons", JINR Communic. Ð3-81-828, Dubna, 1981 (in russian).

[26] M.I. Novopoltsev, Yu.N. Pokotilovskii, "Activation experiments with
ultracold neutrons. Measurement of the absorption coefficient for UCN at
subbarrier reflection from copper surface", JINR Communic. Ð3-85-843, Dubna,
1985 (in russian).

M.I. Novopoltsev, Y. Panin, Y.N. Pokotilovskii, E.V.Rogov, V.A.Stepanchikov,
and I. G. Shelkova, Z.Phys. B 70 (1988) 199-202.

[27] M.I. Novopoltsev, Yu.N. Pokotilovskii, "Measurement of the ultracold neutron
reflection from the surface of substances with large capture cross section",
JINR Communic. Ð3-87-408, Äóáíà, 1987 (in russian).

V.I. Morozov, M.I. Novopoltsev, Y. Panin, Y.N. Pokotilovskii, E.V.Rogov,
JETP Lett., 46 (1987) 377.

[28] M.I. Novopoltsev, Yu.N. Pokotilovskii, "Shuttering of the UCN flux by
magnetized ferromagnetic films", JINR Communic. 3-11986, Dubna,
1978 (in russian).

[29] M.I. Novopoltsev, Yu.N. Pokotilovskii and I. G. Shelkova,
Nucl. Instr. Meth., A264 (1988) 518.

\newpage

\begin{figure}
\begin{center}
\resizebox{18cm}{22cm}{\includegraphics[width=\columnwidth]{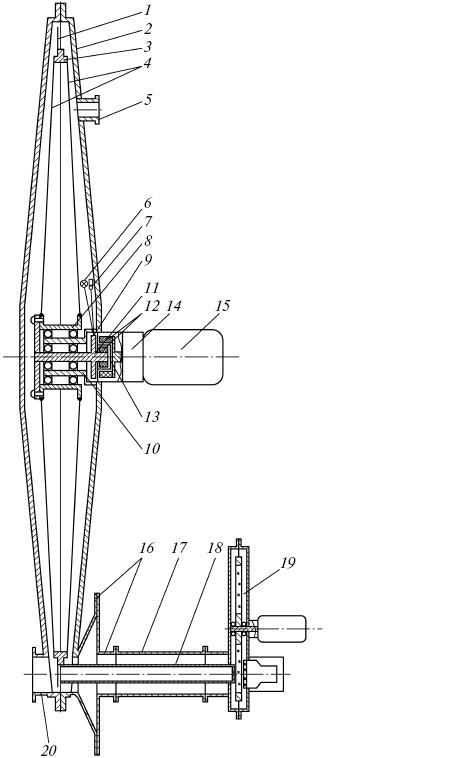}}
\end{center}
\caption{Fig. 1. Scheme of the UCN correlation spectrometer, explanation in the
text.}
\end{figure}

\begin{figure}
\begin{center}
\resizebox{16cm}{18cm}{\includegraphics[width=\columnwidth]{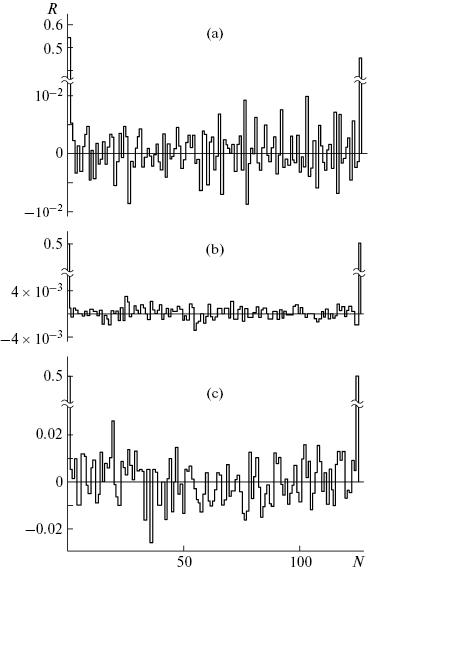}}
\end{center}
\caption{Fig. 2. Theoretical resolution functions of the correlation UCN
spectrometer under assumptions (a) that the operating periods of the chopper
and the analyzer differ by 0.1
the chopper windows is within the limits (b) of 1.25 and (c) 12.5
elementary interval width.}
\end{figure}

\begin{figure}
\begin{center}
\resizebox{16cm}{18cm}{\includegraphics[width=\columnwidth]{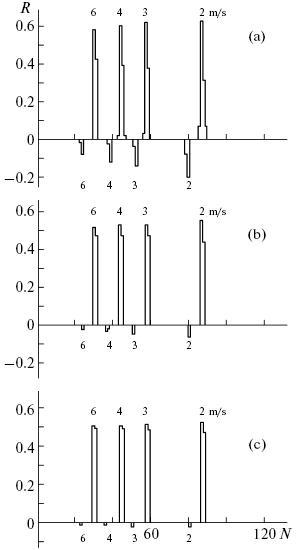}}
\end{center}
\caption{Fig. 3. Model TOF spectra of monoenergetic neutrons with velocities of
2, 3, 4, and 6 m/s at chopper thicknesses (a) of 0.8, (b) 0.2, and (c) 0.1 mm.
 The flight path length was 20 cm, and the duration of the elementary interval
was 1 ms.}
\end{figure}

\begin{figure}
\begin{center}
\resizebox{16cm}{18cm}{\includegraphics[width=\columnwidth]{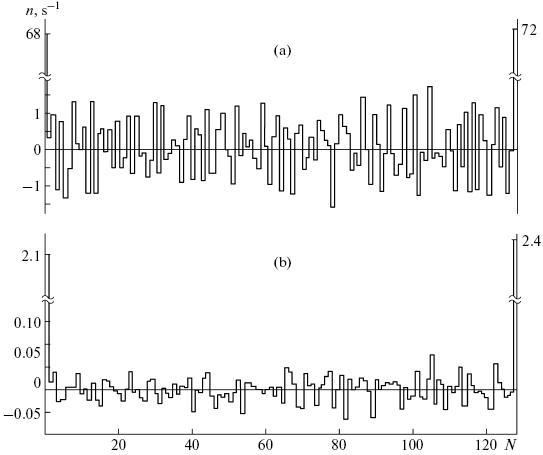}}
\end{center}
\caption{Fig. 4. Resolution functions (a) of the spectrometer and (b) its model,
measured using the alpha-source.}
\end{figure}

\begin{figure}
\begin{center}
\resizebox{16cm}{18cm}{\includegraphics[width=\columnwidth]{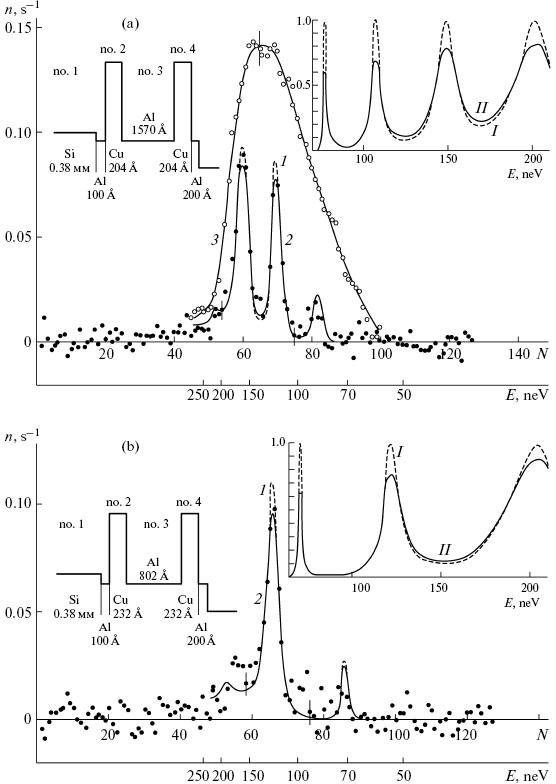}}
\end{center}
\caption{Fig. 5. Level circuit diagram of the interference filter; the inset
presents the calculated neutron transmission factor for (I) the ideal filter
and (II) the Gaussian spread in the thickness of layer 3 with (a) $\sigma$=42
and (b) 28 \AA; the experimental spectrum (1) for the ideal filter and (2)
the Gaussian spread in the thickness of layer 3 with (a) $\sigma$=42 and (b)
28 \AA is at the center; and (3) spectrum of neutron after passing through the
silicon wafer 0.38 mm thick. The flight path length is 31 cm.}
\end{figure}

\end{document}